\documentclass{desyproc}

\newcommand{\apjs}{Astrophys. J. Suppl. Ser.}
\newcommand{\aap}{Astron. Astrophys.}

\newcommand{\mnras}{Mon. Not. R. Astron. Soc.}
\newcommand{\araa}{Ann. Rev. Astron. Astrophys.}

\newcommand{\jcap}{JCAP}

\newcommand{\prd}{Phys. Rev. D}
\newcommand{\apj}{Astrophys. J.}

\begin{document}
\title{Revisiting the Indication for a low opacity Universe for very high energy $\gamma$-rays}

\author{{\slshape Manuel Meyer, Dieter Horns and Martin Raue}\\[1ex]
Institut f\"ur Experimentalphysik, Universit\"at Hamburg, Luruper Chaussee 149, 22761 Hamburg, Germany}


\desyproc{DESY-PROC-2012-04}
\acronym{Patras 2012} 
\doi  

\maketitle

\begin{abstract}
Very high energy (VHE, energy $\gtrsim 100\,$GeV) $\gamma$-rays undergo pair production with photons of the extragalactic background light (EBL).
Thus, the intrinsic $\gamma$-ray flux of cosmological sources is attenuated and
the Universe should be opaque to $\gamma$-rays above a redshift dependent energy.
Recently, an indication has been found that the Universe is more transparent than predicted by a lower-limit EBL model.
Here, this indication is confronted with additional VHE $\gamma$-ray spectra and different EBL models.
Depending on the model for the opacity, the indication persist between a $\sim 2.6\,\sigma$ and $\sim4.3\,\sigma$ confidence level.
\end{abstract}

\section{Introduction}
The Universe is opaque for very high energy (VHE, energy $\gtrsim 100$\,GeV) $\gamma$-rays  
originating from cosmological sources above a certain energy. 
The reason is the interaction of VHE $\gamma$-rays with low energy photons of the extragalactic background light (EBL),
$\gamma_\mathrm{VHE}\,\gamma_\mathrm{EBL} \to e^{+}e^{-}$ \cite{gould1966}.
The EBL ranges from  optical / ultra-violet (UV) to far infrared wavelengths and has its origin in the emission of stars integrated 
over the history of the Universe as well as star emission absorbed and re-emitted by interstellar dust in galaxies \cite[for a review]{hauser2001}.
As a consequence, the intrinsic flux of a VHE $\gamma$-ray source\footnote{Most known extragalactic VHE $\gamma$-ray sources are blazars, i.e. active galactic nuclei (AGN)
with the jet of relativistic particles pointed along the line of sight to the observer.}
is dimmed exponentially by $\exp(-\tau(z,E))$, where $\tau$ is the so called optical depth, an increasing function with the energy $E$,
the redshift $z$ of the source, and the EBL photon density. 

Prominent foreground emission makes a direct observation of the EBL difficult \cite{hauser1998}, so one has to rely on upper limits derived from VHE $\gamma$-ray
observations \cite[for recent upper limits]{meyer2012}, 
lower limits in the optical and infrared inferred from galaxy number counts \cite{madau2000, fazio2004}, or EBL models \cite[as examples]{franceschini2008,kneiske2010,dominguez2011}.
The EBL model of \cite[henceforth KD EBL]{kneiske2010} closely follows the lower limits in the infrared and thus predicts a minimal attenuation for VHE $\gamma$-rays.
It has been used in \cite[henceforth HM12]{horns2012} to test if the Universe is even more transparent and indeed an indication for a \emph{pair production anomaly} (PPA) was found.
The authors analyze a large sample of VHE $\gamma$-ray spectra in the following way. 
All data points of each intrinsic spectrum are fitted with analytical functions and the residuals are calculated. 
For a reasonable fit to the spectra, the residuals $\chi$, defined as the differece between the model flux and the observed flux normalized 
to the statistical uncertainty on the flux (the latter two are absorption corrected) should scatter around a mean of zero.
If the KD EBL correctly describes the attenuation, than this should also be true for the optical thick regime (i.e. $\tau \ge 2$) and this hypothesis is tested with a Student's $t$-test 
(see HM12 for further details).
With this method, HM12 find a $\sim 4\,\sigma$ indication that the KD EBL model over-corrects the observed data and thus the opacity seems to be over-estimated.

For the present analysis, all spectral listed in Tab. 1 of HM12 are considered together with the five extra spectra listed in Tab. \ref{tab:spectra}.
Only the spectrum of 3C\,279 has one data point in the optical thick regime for the KD EBL.
All other spectra only contribute to the $(\tau \ge 2)$-distribution if higher EBL density models are considered.
Besides the minimal attenuation KD EBL, the models of \cite[henceforth FRV EBL]{franceschini2008} and \cite{dominguez2011} are used to correct the VHE $\gamma$-ray spectra
for absorption. 
The latter two models predict a very similar attenuation and a low EBL density at UV / optical wavelengths 
resulting in a lower opacity for energies $E \lesssim 300$\,GeV than the KD EBL.
For higher energies, the models result in a stronger attenuation. 
The effect of these two models on the significance of the PPA is, however, non-trivial as it also depends on the actual shape of the VHE spectra.
It is also allowed for an additional scaling of the optical depth by a factor of 1.3.
This value is motivated by the detection of the EBL in VHE $\gamma$-ray spectra with a best-fit scaling factor of $\sim 1.3$ for the optical depth \cite{hess2013ebl}.
The scaled and unscaled EBL models also frame the EBL density allowed by the lower and upper limits.

\begin{table}[thb]
 \centering
\begin{small}
 \begin{tabular}{|lcccc|}
  \hline
   Source & Redshift & Experiment & Energy range (TeV)&  Reference\\
  \hline
   B3\,2247+381 & 0.119 & MAGIC & 0.24 -- 0.93 & \cite{rxj0648veritas2011} \\
   RX\,J0648.7+1516 & 0.179 & VERITAS & 0.21 -- 0.48 & \cite{b32247magic2012}\\
   RBS\,0413 & 0.190 & VERITAS& 0.23 -- 0.61  & \cite{rbs0413veritas2012}\\ 
  1ES\,0414+009& 0.287 & VERITAS & 0.23 -- 0.61 & \cite{1es0414veritas2012}\\
   3C\,279 & 0.536 & MAGIC & 0.15 -- 0.35  & \cite{3c279magic2011}\\
\hline
 \end{tabular}
\end{small}
\caption{List of additional VHE $\gamma$-ray spectra added to the analysis conducted in HM12. For all other spectra and the corresponding references see HM12, Tab. 1.}
\label{tab:spectra}
\end{table}

\section{Results}
\begin{figure}[t]
 \centering
 \includegraphics[width = 0.85\linewidth]{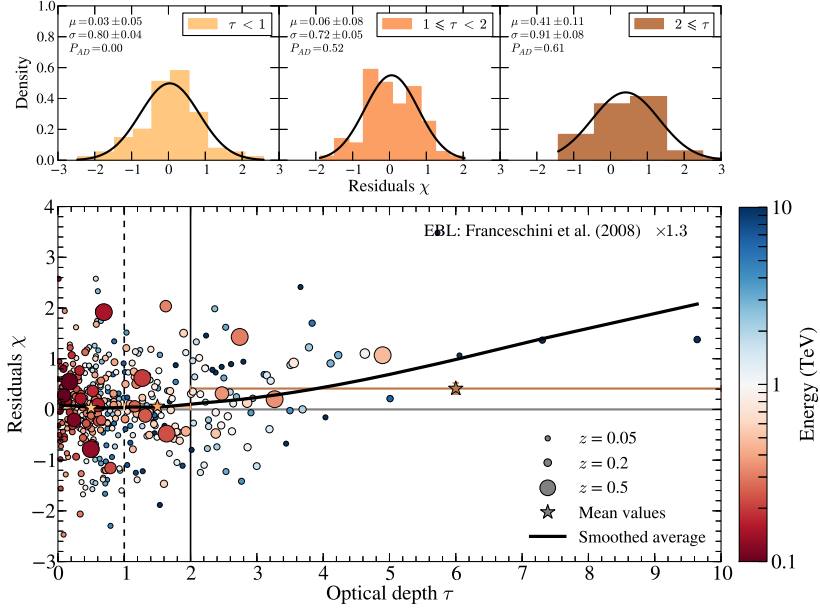}
\caption{Distribution of the fit residuals in the different regimes of the optical depth for the FRV EBL. Top panel: histograms of the residual distribution for
the regimes (from left to right) $\tau < 1$, $1 \le \tau < 2$ and $2 \le \tau$. The mean, variance, and probability of a Gaussian fit are also shown.
Bottom panel: residuals as a function of the optical depth. The marker sizes correspond to the redshift of the objects and the color coding represents the 
energies of each data point. The black solid line shows a smoothed average and the stars mark the mean of the distributions.
}
\label{fig:results}
\end{figure}

The results of the significances of the $t$-test for the PPA are shown in Tab. \ref{tab:results}.
In general, the models predicting a higher EBL density result in lower significances.
The reason is the migration of residuals 
into the optical thick regime which broadens the distribution and shifts it closer to zero (cf. histograms in the top panel in Fig. \ref{fig:results})
However, the overall trend of the residual distribution is unchanged (lower panel of Fig. \ref{fig:results}). 
The black solid line shows a smoothed average\footnote{It is based on a locally weighted polynomial regression (LOESS) \cite{cleveland1979}} of the residual distribution. 
For increasing values of $\tau$ the average also increases towards positive values of the residuals. 
The right most columns of Tab. \ref{tab:results} show the results if only spectra are included that have data points in the optical thick regime in KD EBL.
For all EBL models, these are the spectra with the highest $\tau$ measurements and all significances increase for a unity scaling of $\tau$.
This underlines that the indication of the PPA is especially due to high optical depth measurements.

The smallest significance, i.e. the least tension between EBL model and data, is found for the KD EBL scaled by 1.3.
The significance reduces to just above $2\,\sigma$. 
However, the high ultra-violet / optical density in this scaled model is difficult to reconcile with recent measurements 
of the optical depth with the \emph{Fermi}-LAT \cite{ackermann2012}. 
Using a sample of 150 BL Lac objects up to a redshift of 1.6, the authors of \cite{ackermann2012} obtain a best-fit value of the scaling of the optical depth of $0.90\pm0.19$ for the KD EBL.

\begin{table}[thb]
 \centering
\begin{small}
 \begin{tabular}{|lc|cc|cc|}
\hline
{} & {} & {All} & {spectra} & {Only} & {$\tau_\mathrm{KD}\ge 2$ spectra} \\
EBL model &{} &  $\tau \times 1.0$ & $\tau \times 1.3$ & $\tau \times 1.0$ & $\tau \times 1.3$  \\
\hline
Franceschini et al. (2008, FRV EBL) & \cite{franceschini2008} & 2.60 & 3.50 & 3.45 & 2.92 \\
Kneiske \& Dole (2010, KD EBL) & \cite{kneiske2010} & 4.34 & 2.04 & 4.34 & 2.61 \\
Dominguez et al. (2011) & \cite{dominguez2011} & 2.93 & 3.51 & 3.89 & 3.05 \\
\hline
 \end{tabular}
\end{small}
\caption{Significances (one-sided confidence intervals) of the PPA for the Students $t$-test for different EBL models including either all spectra or only those
with data points corresponding to $\tau\ge2$ in the KD EBL ($\tau = \tau_\mathrm{KD}$).}
\label{tab:results}
\end{table}

\section{Conclusion}
The indication of the pair production anomaly persist with the inclusion of recent VHE $\gamma$-ray spectra in the sample used in HM12 at a $\sim 4.3\,\sigma$ level for the KD EBL.
Different EBL models that predict a higher EBL density can reduce the significance which is due to the migration of data points to the optical thick regime. 
This shows the natural limitation of the method used here: as long as the fit to the spectra is reasonable, choosing a model with 
a very high EBL density will shift most data points into the optical thick regime and the mean of the distribution will be close to zero. 
At the moment, the bulk of VHE emitting blazars has a redshift between 0.1 and 0.2. Thus, a change in the EBL density at infrared wavelengths
has the strongest impact on the results.
In order to give a definite answer on the significance of the PPA, more observations will be necessary, namely of distant objects at several hundreds of GeV, probing the ultra-violet / optical regime 
of the EBL and of sources beyond several tens of TeV to probe the far infrared part of the EBL.
A similar hint for a low opacity Universe has also been found in \emph{Fermi}-LAT data \cite{meyer2012fermi}.

If the PPA persist, it can be hint for a mechanism that reduces the opacity of the Universe. 
For instance, cosmic rays originating from AGN could also interact with the EBL and produce secondary VHE $\gamma$-rays that are responsible
for the highest energies in VHE spectra \cite{essey2010}.
In such scenarios, it is, however, difficult to account for the short term variability in the optical thick regime.\footnote{
The cascade mechanism is especially sensitive to short term variability of distant sources at $z \gtrsim 0.2$ at energies $E\gtrsim1\,$TeV
which has not been observed so far \cite{prosekin2012}. 
On the other hand, for nearby sources such as Markarian\,421 or Markarian\,501,
a sizeable amount of cascade emission can only be produced by protons of energies $E_p \gtrsim 10^{19}\,$eV \cite{aharonian2012} and 
a short-time variability at energies beyond tens of TeV (i.e. $\tau\gtrsim 1$) would not constrain such scenarios.
However, another mechanism would be required to explain how $\gamma$-ray photons can escape pair production in such cases.
}
Another possibility is the oscillations of photons into axion-like particles (ALPs) in ambient magnetic fields.
For example, it has been shown that the ALP production in intra-cluster magnetic fields followed by a reconversion in the magnetic field of the Milky Way
can lead to a substantial hardening of VHE spectra \cite{horns2012ICM}.

\paragraph{Acknowledgments} MM would like to thank the state excellence cluster ``Connecting Particles with the Cosmos'' at the university of Hamburg.


\begin{footnotesize}

\providecommand{\newblock}{}

\end{footnotesize}


\end{document}